\documentclass[twocolumn]{aastex62}

\usepackage{hyperref}
\usepackage{subfigure}
\usepackage{booktabs}
\newcommand{\oi}{\ion{O}{1}}
\newcommand{\oii}{\ion{O}{2}}
\newcommand{\oiii}{\ion{O}{3}}
\newcommand{\nei}{\ion{Ne}{1}}
\newcommand{\neii}{\ion{Ne}{2}}
\newcommand{\neiii}{\ion{Ne}{3}}
\newcommand{\nvi}{\ion{N}{6}}
\newcommand{\nvii}{\ion{N}{7}}
\newcommand{\ovii}{\ion{O}{7}}
\newcommand{\oviii}{\ion{O}{8}}
\newcommand{\neix}{\ion{Ne}{9}}
\newcommand{\nex}{\ion{Ne}{10}}
\newcommand{\nvin}{\ion{N}{6}}
\newcommand{\nviin}{\ion{N}{7}}
\newcommand{\oviin}{\ion{O}{7}}
\newcommand{\oviiin}{\ion{O}{8}}
\newcommand{\neixn}{\ion{Ne}{9}}
\newcommand{\nexn}{\ion{Ne}{10}}
\newcommand{\fexvin}{\ion{Fe}{16}}
\newcommand{\fexviin}{\ion{Fe}{17}}
\newcommand{\fexviiin}{\ion{Fe}{18}}
\newcommand{\fexvii}{\ion{Fe}{17}}
\newcommand{\fexxi}{\ion{Fe}{21}}

\def\nvi{{{\rm N}\,{\sc vi}~}}
\def\nvii{{{\rm N}\,{\sc vii}~}}
\def\oi{{{\rm O}\,{\sc i}}}
\def\oii{{{\rm O}\,{\sc ii}}}
\def\oiii{{{\rm O}\,{\sc iii}}}
\def\ovii{{{\rm O}\,{\sc vii}~}}
\def\oviii{{{\rm O}\,{\sc viii}~}}
\def\neix{{{\rm Ne}\,{\sc ix}~}}
\def\nex{{{\rm Ne}\,{\sc x}~}}
\def\nvin{{{\rm N}\,{\sc vi}}}
\def\nviin{{{\rm N}\,{\sc vii}}}
\def\oviin{{{\rm O}\,{\sc vii}}}
\def\oviiin{{{\rm O}\,{\sc viii}}}
\def\nei{{{\rm Ne}\,{\sc i}}}
\def\neii{{{\rm Ne}\,{\sc ii}}}
\def\neiii{{{\rm Ne}\,{\sc iii}}}
\def\neixn{{{\rm Ne}\,{\sc ix}}}
\def\nexn{{{\rm Ne}\,{\sc x}}}
\def\fexvin{{{\rm Fe}\,{\sc xvi}}}
\def\fexviin{{{\rm Fe}\,{\sc xvii}}}
\def\fexvii{{{\rm Fe}\,{\sc xvii}~}}
\def\fexxi{{{\rm Fe}\,{\sc xxi}~}}
\def\fexviiin{{{\rm Fe}\,{\sc xviii}}}
\def\fexxiv{{{\rm Fe}\,{\sc xxiv}~}}

\def\xmm{{\it XMM-Newton}~}
\def\chandra{{\it Chandra}~}

\shorttitle{Multi-thermal hot CGM with non-solar composition}
\shortauthors{Das et al.}

\begin{document}

\title{Discovery of a very hot phase of the Milky Way CGM with non-solar abundance ratios}
\correspondingauthor{Sanskriti Das}
\email{das.244@buckeyemail.osu.edu}

\author[0000-0002-9069-7061]{Sanskriti Das}
\affiliation{Department of Astronomy, The Ohio State University, 140 West 18th Avenue, Columbus, OH 43210, USA}

\author{Smita Mathur}
\affiliation{Department of Astronomy, The Ohio State University, 140 West 18th Avenue, Columbus, OH 43210, USA}
\affil{Center for Cosmology and Astroparticle Physics, 191 West Woodruff Avenue, Columbus, OH 43210, USA}

\author{Fabrizio Nicastro}
\affiliation{Observatorio Astronomico di Roma - INAF, Via di Frascati 33, 1-00040 Monte Porzio Catone, RM, Italy}
\affiliation{Harvard-Smithsonian Center for Astrophysics, 60 Garden St., MS-04, Cambridge, MA 02138, USA}
\author{Yair Krongold}
\affiliation{Instituto de Astronomia, Universidad Nacional Autonoma de Mexico, 04510 Mexico City, Mexico}

\begin{abstract}
\noindent We present the discovery of a very hot gas phase of the Milky Way circumgalactic medium (CGM) at T $\approx 10^7$ K, using deep \xmm RGS observations of the blazar 1ES\,1553+113. The hot gas, coexisting with a warm-hot phase at T $\approx 10^6$ K is $\alpha-$enhanced, with [O/Fe] = 0.9$^{+0.7}_{-0.3}$, indicating core-collapse supernovae enrichment. Additionally we find [Ne/O] and [N/O] = $0.7^{+1.6}_{-0.2}$, such that N/Ne is consistent with solar. Along with the enrichment by AGB stars and core-collapse supernovae, this indicates that some oxygen has depleted onto dust and/or transited to cooler gas phase(s). These results may affect previous baryonic and metallic mass estimations of the warm-hot and hot CGM from the observations of oxygen emission and absorption. Our results provide insights on the heating, mixing and chemical enrichment of the Milky Way CGM, and provide inputs to theoretical models of galaxy evolution. 
\end{abstract}


\section{Introduction} \label{sec:intro}
\noindent The circumgalactic medium (CGM) is the halo of multi-phase gas and dust surrounding the stellar component and interstellar medium (ISM) of galaxies, inside their virial radii \citep{Tumlinson2017}. It is a very important component of a galaxy harboring a large fraction of its missing baryons and missing metals \citep{Gupta2012,Peeples2014}. Numerical simulations show that the properties of the CGM are governed by halo mass, and are affected by accretion from the intergalactic medium (IGM) and feedback from the galactic disk \citep{Ford2014,Suresh2017,Oppenheimer2018}. Precipitation from the CGM in turn may help sustain next generation of star-formation in a galaxy \citep{Voit2015}. Thus the CGM plays an instrumental role in the evolution of a galaxy. 

The CGM is multiphase in its ionization states, spanning 2 orders of magnitude of temperature: T $\approx$ 10$^{4-6}$ K \citep{Ford2014,Suresh2017}. In this paper, we focus on the hot ($\geqslant$10$^6$ K) CGM of Milky Way. The hot (T $\approx$ 10$^6$ K) gaseous Galactic corona at the virial temperature is a long-standing prediction \citep{Spitzer1956}. For halos $\geqslant$ 10$^{12}$ M$_\odot$, cold rarefied and metal-poor infalling gas is shock-heated to the virial temperature as it enters the halo. On the other hand, dense metal-enriched galactic outflow driven by the winds of massive stars, supernovae and AGN (active galactic nuclei) feedback contribute to the hot CGM as well. The quasi-static galactic corona is in general a mix of both, and is not necessarily mono-thermal and of solar-like chemical composition. Therefore, studying the abundances in the highly ionized CGM and its different thermal components, if any, are extremely important to understand the thermodynamics, mixing and chemical evolution of the CGM. Deep X-ray absorption spectroscopy, where the hot CGM can be probed by He-like and H-like ionized metals, is a great tool for this. 

Because of our special vantage point, the warm-hot CGM of Milky Way has been studied via emission and absorption in much better details compared to other galaxies. It has been observed to be diffuse, extended, massive and anisotrpic \citep{Henley2010,Gupta2012,Henley2013,Nicastro2016b,Nakashima2018}. Absorption studies show that there is a range of \ovii column densities along different sightlines across the sky, but the temperature is similar, about $10^6$ K \citep[e.g.][]{Gupta2012}. Similarly, emission studies show that while the emission measure varies by an order of magnitude across the sky, the temperature of the CGM is practically constant \citep{Henley2013}. Thus, from both the emission and absorption studies, the diffuse warm-hot CGM was believed to be of single temperature. There were hints of hotter components in previous emission studies, but these were questionable due to confusion with the foreground components \citep{Henley2013,Nakashima2018}.  In this paper, we present a deep 1.85 Ms \xmm RGS (Reflection Grating Spectrometer) observation of blazar 1ES\,1553+113; the sightline probes the Milky Way CGM in absorption with unprecedented sensitivity. Detailed analysis and spectral modelling of the data has led to the discovery of the hottest component of Milky Way CGM at T $\approx 10^7$ K, coexisting with a warm-hot phase at T $\approx 10^6$ K. We also find that the hot gas is $\alpha-$enhanced, indicating core-collapse supernovae enrichment. Additionally, we find non-solar abundance ratios of N, O and Ne in the warm-hot phase. 

Our paper is structures as follows. We discuss data analysis in section \ref{sec:analysis}, and results in section \ref{sec:results}. We interpret our results and discuss their implications in section \ref{sec:discussion}. Finally, we summarize our results and outline the future aim in section \ref{sec:summary}.

\section{Observation and analysis}\label{sec:analysis}
\begin{figure}
\centering
\includegraphics[trim=15 0 72 31, clip, scale=0.33]{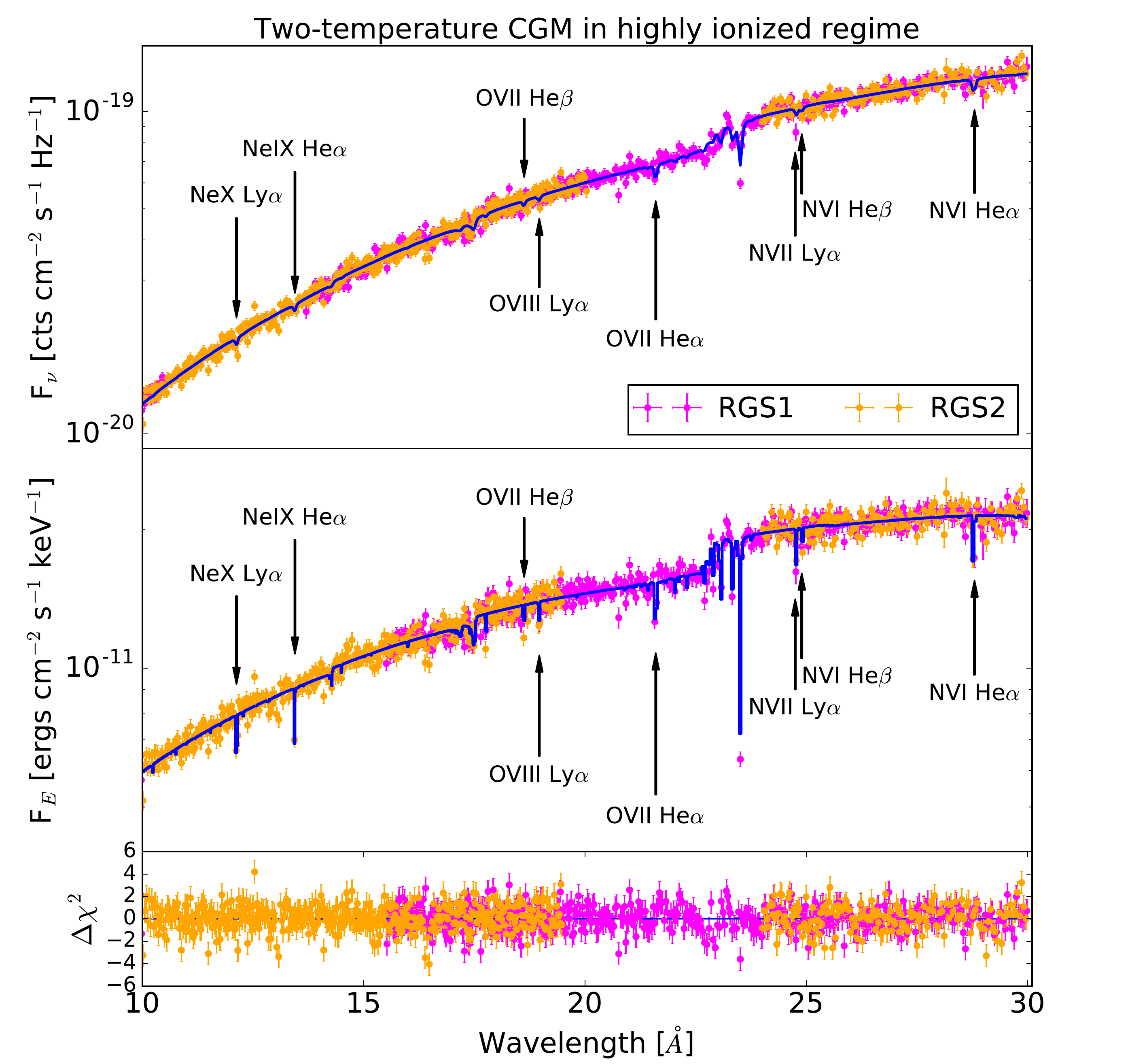}
\caption{\label{fig:spectrum}The data (top) and unfolded spectra (middle), in bins with a SNR per bin $\geq 25$ of 1ES\,1553+113 and the best-fitting model with two \texttt{PHASE} components (blue). The residuals are plotted in the bottom panel.}
\end{figure}

\begin{table*}
\centering
\caption{Absorption line parameters. Uncertainties are represented as 1$\sigma$ errors. The second last column is the summation of the 5th and 6th columns, which are the results of \texttt{PHASE} modelling. The equivalent widths and corresponding column densities obtained from Gaussian line fitting are in the 4th and the last column, respectively. The column densities of the $\alpha$ and $\beta$ transitions of \nvi and \ovii have been shown separately in the last column. } \label{tab:lineprofiles}
\begin{tabular}{c c c c c c c c}
\hline
\hline 
 Ion & Transition & $\lambda$ &  EW & N$_{T_1}$ & N$_{T_2}$ & N$_{tot}$ & N(from EW) \\
 & & (\AA) & (m\AA) & $(cm^{-2})$ & $(cm^{-2})$ & $(\times 10^{15} cm^{-2})$ & $(\times 10^{15} cm^{-2})$ \\
\hline 
\nvi &  &  & & 4.4$^{+1.5}_{-1.3}\times 10^{15}$  & 4.8$^{+3.2}_{-3.1}\times 10^{12}$ & 4.4$^{+1.5}_{-1.3}$ & 9.2$\pm$2.8 \\
& He$\alpha$ & 28.787 & 9.5$_{-2.8}^{+3.1}$ &   &  &  & 1.9$\pm0.6$\\ 
& He$\beta$ & 24.898 & 5.7$\pm$2.2 &  &  &  & 7.3$\pm$2.8\\ 
\nvii & Ly$\alpha$ & 24.781 & 5.7$\pm$2.5 & 2.3$^{+0.8}_{-0.7} \times 10^{15}$  & 6.8$^{+4.6}_{-4.5}\times 10^{14}$ & 3.0$_{-0.8}^{+0.9}$ & 2.5$\pm$1.1\\ 
\ovii &  &  & & 9.3$_{-3.1}^{+3.5}\times 10^{15}$  & 4.8$_{-3.2}^{+3.3}\times 10^{13}$ & 9.3$_{-3.1}^{+3.5}$ & 11.6$\pm$5.4\\ 
& He$\alpha$ & 21.602 & 10.6$\pm2.8$ &   &  &  & 3.7$\pm1.0$\\
& He$\beta$ & 18.627 & 3.6$\pm$2.4 &  &  &  & 7.9$\pm$5.4\\ 
\oviii & Ly$\alpha$ & 18.967 & 4.5$\pm1.9$ & 5.3$^{+2.0}_{-1.8}\times 10^{14}$  & 2.7$^{+1.9}_{-1.8}\times 10^{15}$ & 3.2$^{+1.9}_{-1.8}$ & 3.4$\pm1.4$\\ 
\neix & He$\alpha$ & 13.447 & 6.3$_{-2.7}^{+2.9}$ & 7.9$^{+4.5}_{-3.4}\times 10^{15}$  & 1.4$^{+0.7}_{-0.6}\times 10^{15}$ & 9.3$^{+4.7}_{-3.4}$ & 5.4$\pm2.3$\\ 
\nex & Ly$\alpha$ & 12.134 & 8.5$_{-2.6}^{+2.7}$ & 6.0$^{+3.5}_{-2.6}\times 10^{12}$  & 13.7$^{+6.5}_{-5.9}\times 10^{15}$ & 13.7$^{+6.5}_{-5.9}$ & 15.8$\pm4.6$ \\ 
\hline
\end{tabular} 
\end{table*} 
\noindent Our target, 1ES\,1553+113, was observed by \xmm for 1.85Ms\footnote{\xmm ObsIDs: 0094380801, 0656990101, 0727780101, 0727780201, 0727780301, 0761100101, 
0761100201, 0761100301, 0761100401, 0761100701, 0761110101, 0790380501, 0790380601, 0790380801, 0790380901, 0790381001, 0790381401, 0790381501.}. This deep observation yielded high signal-to-noise ratio per resolution element (SNRE$\approx43$) X-ray grating spectra, with RGS1 and RGS2, presented in figure \ref{fig:spectrum}. The details of the observations and data reduction are presented in \cite{Nicastro2018}, discussing the intervening warm-hot intergalactic medium (WHIM). Here we focus on the $z=0$ absorption lines in the 8--29 \AA\ range, probing the circumgalactic medium (CGM) of the Milky Way. Strong lines of highly ionized metals at $z=0 $ are clearly seen at the expected wavelengths (Table \ref{tab:lineprofiles}). All the spectral analysis is performed using \texttt{XSPEC}\footnote{\url{https://heasarc.gsfc.nasa.gov/xanadu/xspec/manual/XspecManual.html}}. Because a \ovii He$\beta$ WHIM line was detected around 26.5 \AA\citep{Nicastro2018}, we remove the channels in the 26--27 \AA~ range while fitting the data. Also, there are absorption-like features around 10 \AA, 13 \AA, 16.5 \AA, 17.1 \AA, 17.9 \AA, 20.5 \AA, 28 \AA~ (figure \ref{fig:spectrum}), which are likely the instrumental features due to cool pixels in the detectors. We have fit the data twice by keeping and removing the channels around these wavelengths; the continuum is similar either way. However, the $\chi^2$ improves significantly in the absence of these channels.

\subsection{$z=0$ lines and model-independent results}\label{sec:identify}
\noindent To detect and characterize the absorption lines, we need to determine the source continuum first, which we model as a power-law plus blackbody spectrum, absorbed by the cold and warm ISM of the Milky Way (model \texttt{ISMabs*(powerlaw+bbody)} in \texttt{XSPEC}). \texttt{ISMabs} accounts for the neutral, singly and doubly ionized metal lines and takes care of the oxygen and neon edges \citep{Gatuzz2016}. We allowed the column densities of H, O (\oi, \oii, \oiii), Ne (\nei, \neii, \neiii) and Fe to vary; other elements do not affect the spectra significantly, so we fix their column densities at the relative solar abundances. This resulted in good fit of the continuum, but absorption lines from highly ionized elements were clearly seen in the residuals. We modelled these lines with Gaussian profiles with central wavelengths fixed at the known $z=0$ values of respective ions. We detect the \nvi He$\alpha$ and He$\beta$, \nvii Ly$\alpha$, \ovii He$\alpha$ and He$\beta$, \oviii Ly$\alpha$, \neix He$\alpha$ and \nex Ly$\alpha$ lines at 3.4$\sigma$, 2.6$\sigma$, 2.3$\sigma$, 3.9$\sigma$, 1.5$\sigma$, 2.4$\sigma$, 2.4$\sigma$ and 3.4$\sigma$ significance respectively\footnote{we define the single-line statistical significance as EW/$\Delta$(EW). The effective-area corrections add $\approx$1.2 m\AA~ systematic uncertainty to the statistical uncertainty \citep{Nevalainen2017}.}. We calculate the column density of these lines from their respective equivalent widths assuming that they are in the linear regime of curve of growth (table \ref{tab:lineprofiles}). Also, we repeat the analysis by removing the channels at and around the wavelengths of expected transitions to obtain the continuum, and add the channels back to calculate the equivalent widths. The equivalent width values are similar as those obtained earlier. This confirms that the continuum is not incorrectly estimated due to the presence of absorption lines, and the equivalent widths are not over/underestimated. 

Our detected absorption line strengths suggest several interesting aspects of the observed system. In collisional ionization equilibrium (CIE)\footnote{We assume the gas to be in CIE. The cosmological hydrodynamic simulations \citep{Dave2010} and the high resolution simulations focused on individual galaxies \citep{Stinson2012} show that the CGM is in CIE. Previous X-ray observations \citep{Henley2010,Gupta2012,Henley2013,Gatuzz2018} are also consistent with the plasma being in CIE, validating our assumption.}, the fractional ionization of \ovii peaks \textcolor{black}{($f_{\hbox{\ovii}}\geqslant$0.9)} in the temperature range $10^{5.6-6.2}$ K, and that of \oviii peaks at $\approx10^{6.4}$ K. Similarly, the fractional ionization of \nvi peaks \textcolor{black}{($f_{\hbox{\nvi}}\geqslant$0.9)} at $10^{5.4-6.0}$ K and of \nvii at $\approx10^{6.2}$ K, the fractional ionization of \neix peaks \textcolor{black}{($f_{\hbox{\neix}}\geqslant$0.9)} at $10^{6.1-6.3}$ K and of \nex at $\approx10^{6.6}$ K. The column density ratio of the same element's two ions can uniquely determine the temperature. The \nvi to \nvii ratio implies T $= 10^{5.98-6.09}$ K, \ovii to \oviii ratio implies T $= 10^{6.14-6.26}$ K while the \neix to \nex ratio implies T $= 10^{6.50-6.71}$ K. Clearly, the temperature windows do not overlap with each other. This already suggests that a single temperature model may be inadequate to characterize the observed spectra. Secondly, the column density of neon and nitrogen are as large as that of oxygen: $\frac{\hbox{N(\nvin)+N(\nviin)}}{\hbox{N(\oviin)+N(\oviiin)}}$ = 0.8$\pm$0.4, $\frac{\hbox{N(\neixn)+N(\nexn)}}{\hbox{N(\oviin)+N(\oviiin)}}$ = 1.4$\pm$0.6, even though oxygen is far more abundant than neon and nitrogen in solar composition \citep[$A_{O,\odot} = 4.9\pm0.6 \times 10^{-4}$, $A_{Ne,\odot} = 0.8\pm0.2 \times 10^{-4}$, $A_{N,\odot} = 0.7\pm0.1 \times 10^{-4}$;][]{Asplund2009}. This suggests that the neon and nitrogen are super-solar relative to oxygen in the observed phase. \fexviin--\fexxiv lines and the Fe unresolved transition array (UTAs) of \fexvin--\fexviiin~ are not detected at better than $1\sigma$ significance. This suggests $\alpha$-enhancement relative to iron. 

\subsection{\texttt{PHASE} modelling of $z=0$ absorbers}\label{sec:phase}
\noindent To confirm and quantitatively determine the suggestive results obtained in \S\ref{sec:identify}, we now model the data in detail. We use the hybrid-ionization model \textcolor{black}{(models of collisionally-ionized gas perturbed by photoionization by the meta-galactic radiation field, at a given redshift)} \texttt{PHASE} \citep{Nicastro2018} to fit the data and calculate the temperature, equivalent H column density and the relative abundance of the metals in the Galactic warm-hot/hot absorbers. The model assumes relative abundances and absolute metallicities to be solar by default, but allows them to vary between 0.01 to 100 times solar, independently for each elements from He to Ni (He, C, N, O, Ne, Mg, Al, Si, S, Ar, Ca, Fe and Ni). 

As N, O and Ne have prominent absorption features (figure \ref{fig:spectrum}), and Fe lines (primarily, \fexvii at 12.123\AA, and \fexxi at 12.165\AA) can contaminate the Ne lines through blending, we vary the abundance of N, O, Ne and Fe only. The spectrum is insensitive to the abundance of other elements, so we fix their abundance with respect to oxygen at solar. \textcolor{black}{As noted above, the CGM is assumed to be in CIE; therefore  we fix the photoionization parameter\footnote{\textcolor{black}{Ionization parameter \textit{U} is the flux of ionizing photons per unit density of gas: $U = \frac{Q(H)}{4\pi r^2n(H)c}$, where Q(H) is the number of hydrogen ionizing photons s$^{-1}$, r is the distance to the source, n(H) is the hydrogen density.}} to the lowest possible value allowed by PHASE, at $U=10^{-4}$. This ensures that photo-ionization is negligible and that the gas is collisionally ionized. To begin with, we fit the spectra with a single-temperature \texttt{PHASE} model with solar composition: \texttt{ISMabs*(powerlaw+bbody)*PHASE}. The fit is not very good ($\chi^2/dof$ = 2563.54/2479), showing that the abundance ratios are likely non-solar. Therefore, we fit
the spectra with a single temperature model, but with non-solar composition; this improves the fit ($\chi^2/dof$ = 2548.07/2476), but it cannot reproduce all the detected lines in the spectra; in particular it does not account for the \nex line and it underestimates the \oviii absorption. Therefore we fit the spectra with 
two \texttt{PHASE} models: \texttt{ISMabs*(powerlaw+bbody)*PHASE1*PHASE2}.}
We allow the H column density and the temperatures in two phases to be different, but do not force them to be different. However, we force the relative abundances (N/O, Ne/O, O/Fe) to be same in both phases, which implies similar chemical composition and hence, similar sources of metal enrichment. The best-fitted model ($\chi_\nu^2 = 1.02714$; dof:2471, P($\chi^2$,dof) = 0.622$\times$P$_{max}$) reproduces the lines found in the previous section well (figure \ref{fig:spectrum}, table \ref{tab:lineprofiles})\footnote{\textcolor{black}{The two-temperature solar composition model provides a better fit ($\chi^2/dof$ = 2557.75/2476) than the single-temperature solar composition model, showing that multiple temperatures are necessary. The two-temperature non-solar composition provides an even better fit ($\chi^2/dof$ = 2538.06/2471) than the two-temperature solar composition, indicating that the observed system has multiple temperature components with non-solar abundance ratios.} }. The best-fitted values of the parameters in \texttt{PHASE} models are quoted in table \ref{tab:best-fit}; \textcolor{black}{the errors are quoted at $3\sigma$ intervals, i.e. at $99.73$\% confidence}. 
\begin{table}
\renewcommand{\thetable}{\arabic{table}}
\centering
\caption{Parameters of the best-fitted \texttt{PHASE} model \citep[uncertainties are represented as 99.73\% confidence intervals, and abundances are in log$_{10}$ with respect to solar composition, according to the prescription of][]{Asplund2009}}
\label{tab:best-fit}
\begin{tabular}{c c}
\hline
\hline
Parameters & Values \\
\hline 
log\,T$_1$(K) & $6.11^{+0.19}_{-0.49}$ 
\\ \\ 
log\,T$_2$(K) & $7.06^{+0.80}_{-0.72}$ 
\\ \\ 
$[$Ne/O$]$ & $0.72^{+1.57}_{-0.25}$  \\ \\ 
$[$N/O$]$  & $0.75^{+1.57}_{-0.25}$ \\ \\ 
$[$O/Fe$]$ & $0.86^{+0.67}_{-0.33}$  \\ \\ 
\hline
$[$Ne/Fe$]$ & $1.59^{+1.16}_{-0.38}$  \\ \\ 
$[$N/Fe$]$  & $1.62^{+0.56}_{-0.30}$ \\ 
\hline
\hline 
\end{tabular} 
\end{table} 

The column densities of the detected ions from the two-temperature \texttt{PHASE} model are presented in table \ref{tab:lineprofiles} (5th and 6th column). The sum of these column densities is given in the 7th column. Comparing these with the column densities estimated from the curve-of-growth analysis (last column), we find that the two are in excellent agreement for \ovii and the H-like ions. \nvi and \neix columns are consistent within 2$\sigma$. It should be clarified that we do not use the results of Gaussian line-fitting as a prior in the \texttt{PHASE} modelling. Therefore, fitting the data using \texttt{PHASE} is completely independent from the curve-of-growth analysis in the previous section. Because \texttt{PHASE} takes into account line saturation by Voigt profile fitting, the agreement between the two shows that the effect of saturation in the absorption lines, if any, is negligible.

\section{Results}\label{sec:results}
\noindent The best-fitted \texttt{PHASE} model has yielded 3 interesting results. We have found a very hot gas phase detected with \nex absorption line, coexisting with warm-hot CGM; we have found super-solar O/Fe and super-solar N/O and Ne/O. The parameters of the best-fitted model are presented in table \ref{tab:best-fit}. We quote all uncertainties in 1$\sigma$ error bars, unless explicitly mentioned otherwise. 

\begin{figure}[h]
\centering
\includegraphics[trim=25 0 0 10,clip, scale=0.5]{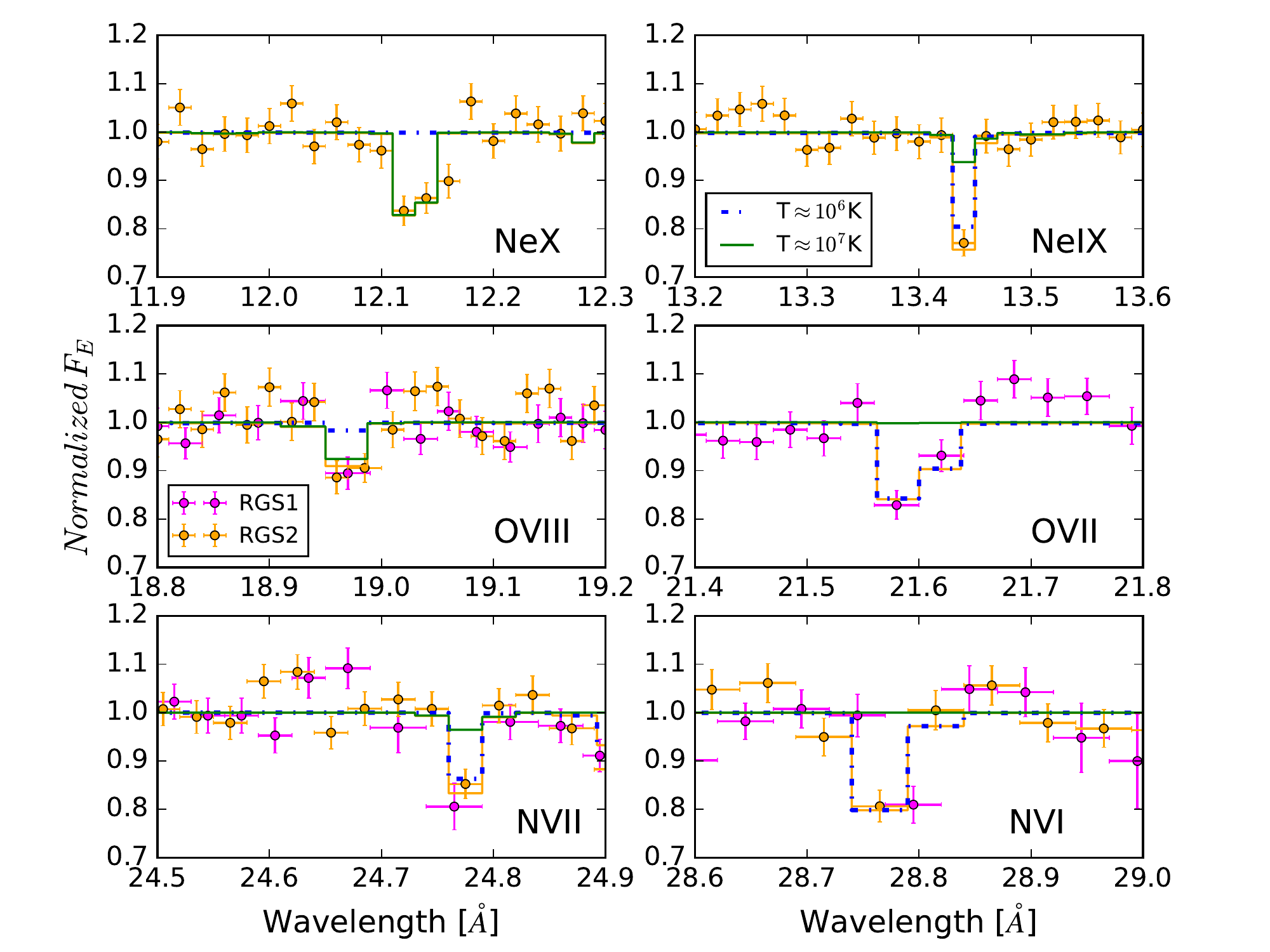}
\caption{\label{fig:components}The absorption by the warm-hot (dotted blue lines) and hot (solid green lines) phases shown separately for each of the detected absorption lines (normalized by the best-fitted continuum model). Magenta points show the RGS1 spectrum and the yellow points show the RGS2 spectrum.}
\end{figure}

\begin{figure*}
\begin{subfigure}
\centering
\includegraphics[trim=10 10 0 10,clip,scale=0.45]{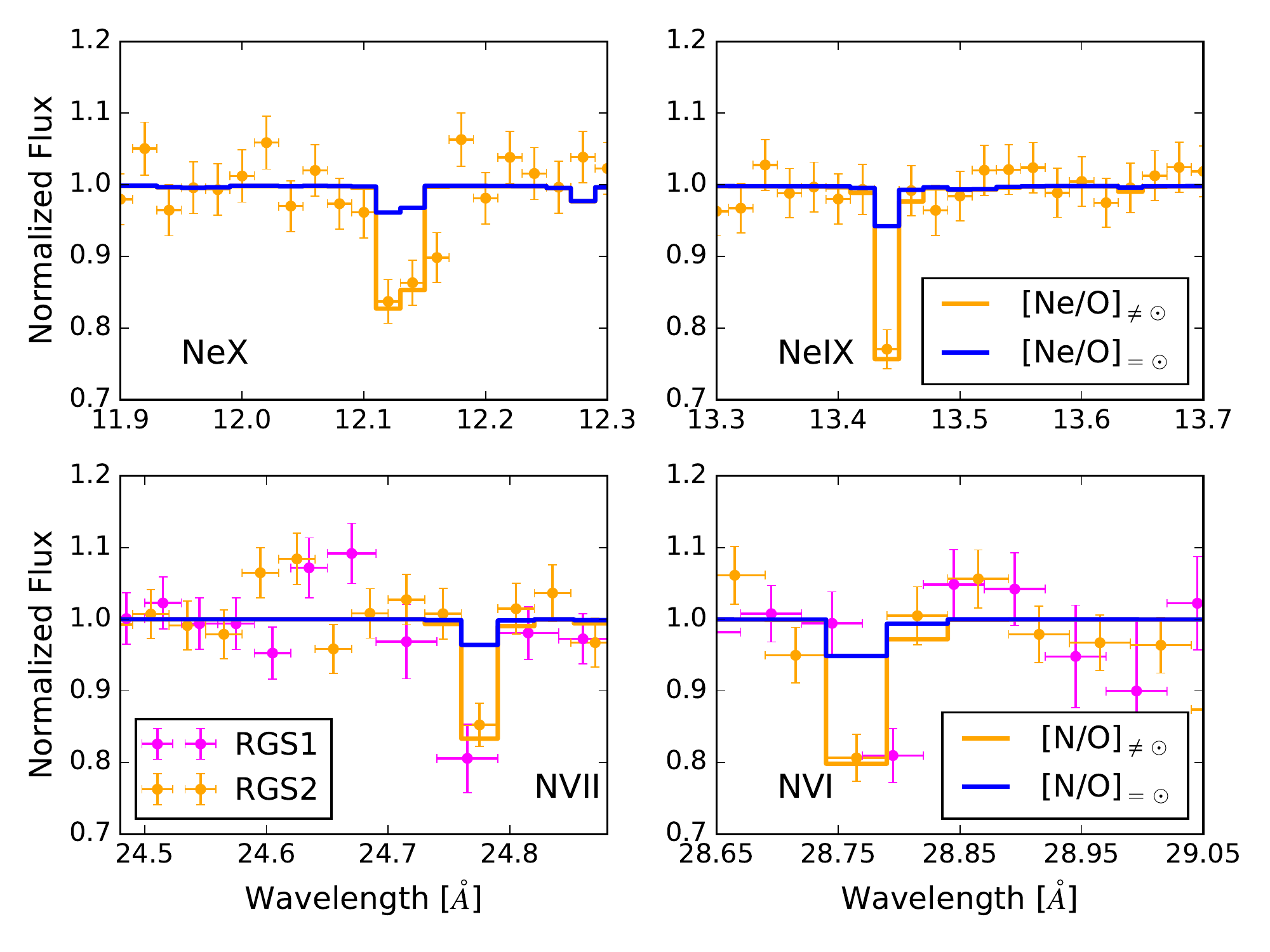}
\end{subfigure}
\begin{subfigure}
\centering
\includegraphics[trim=10 10 0 10, clip,scale=0.45]{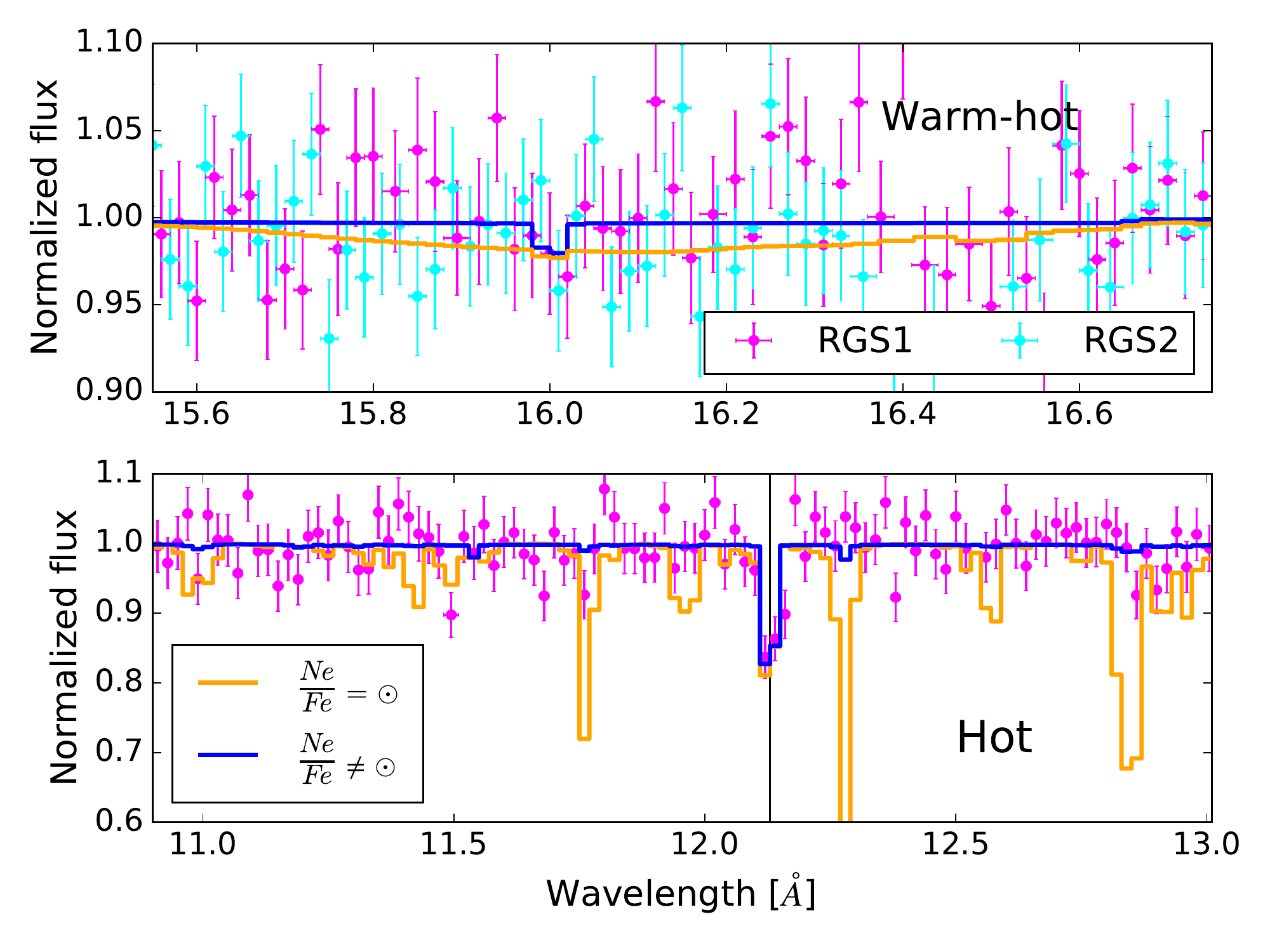}
\end{subfigure}
\caption{\label{fig:nonsolar}Non-solar abundance ratios in the warm-hot and hot CGM. Left: The absorption lines (normalized by the best-fitted continuum model) for the best-fit non-solar  (yellow lines) vs. solar (blue lines) abundance ratios relative to oxygen for the detected Ne and N lines. The observed spectra clearly require non-solar ratios. \label{fig:nonsolarfe} Right: The absorption profile for non-solar [Ne/Fe] vs. the solar [Ne/Fe]. In both the top and bottom panels, the blue lines correspond to the best-fit [Ne/Fe], while the yellow lines are for solar [Ne/Fe], the magenta and cyan points are the data. In the top panel, the broad feature shown by the yellow line  corresponds to the Fe UTA that would have been produced in the warm-hot phase with solar abundances. The weak (blue) line at about 16 \AA~ is from \fexviiin. In the bottom panel, the yellow lines correspond to \fexvii--\fexxiv transitions produced in the hot phase, that would have been detected at the shown strength if [Ne/Fe] was solar. The vertical line in the bottom panel at about 12.1 \AA~ corresponds to the \nex Ly-$\alpha$ line. [Ne/Fe] is clearly super-solar in the hot phase and the warm-hot phase is consistent with the same supersolar best-fit [Ne/Fe].}
\end{figure*}

We find that the temperatures of the two phases are well-constrained and they differ by about an order of magnitude. \nvi is present only at the lower temperature (log (T$_1/K) = 6.1^{+0.2}_{-0.5}$ \textcolor{black}{(3$\sigma$ errors)}, the \textit{warm-hot} phase); \nex is present only at the higher temperature (log (T$_2/K) = 7.1^{+0.8}_{-0.7}$ \textcolor{black}{(3$\sigma$ errors)}, the \textit{hot} phase); \nviin, \oviin, and \neix contribute to both temperature phases, but are predominant at the lower temperature;  and \oviii also contributes to both the phases, but is dominant at the higher temperature (figure \ref{fig:components}). \fexvin--\fexviiin~ UTA and \fexvii to \fexxiv lines are expected to be present at T$_1$ and T$_2$ respectively, but are not detected at a significance of better than $1\sigma$ (figure \ref{fig:nonsolarfe}), resulting a $1\sigma$ upper limit of N(Fe) $<9.1\times10^{15}$ cm$^{-2}$.
As the observed spectral wavelength range contains no lines or edges of hydrogen, we cannot calculate the absolute abundances of metals with respect to hydrogen, or the metallicity. However, we can determine the relative metal abundances. We find that both Ne/O and N/O are super-solar (table \ref{tab:best-fit}) with [Ne/O] as well as [N/O]$=0.7^{+1.6}_{-0.2}$ (\textcolor{black}{3$\sigma$ errors}, figure \ref{fig:nonsolar}). N/Ne is consistent with solar. As noted above, we do not detect absorption lines of iron at better than 1$\sigma$ significance, but using the best fit value and upper limit of iron column density, we can determine the abundance ratios of detected elements relative to iron: [O/Fe]$=0.9^{+0.7}_{-0.3}$; [Ne/Fe]$=1.6^{+1.2}_{-0.4}$; [N/Fe]$=1.6^{+0.6}_{-0.3}$ (\textcolor{black}{3$\sigma$ errors}; table \ref{tab:best-fit}).

The equivalent H column density (modulo metallicity) N$_H^Z$ is uncertain due to the non-solar composition of the gas. Assuming the absolute metallicities of individual elements to be same in both phases, we find that N$_{H,2}^Z$/N$_{H,1}^Z$ is $\approx 22^{+16}_{-13}$. For solar metallicity of oxygen, N$_{H,1}^Z$=$13.3^{+5.0}_{-4.5}\times10^{18}$ cm$^{-2}$, N$_{H,2}^Z$=$2.9^{+2.4}_{-2.0}\times10^{20}$ cm$^{-2}$. 

\section{Discussion}\label{sec:discussion}
\noindent The hot gas phase along with the warm-hot component has not been observed earlier in emission or absorption, so this is the first of its kind. The non-solar abundance ratios in warm-hot CGM is new as well. Below, we discuss the possible origin(s) of the newly discovered hot gas phase. Also, we qualitatively interpret the chemical composition in terms of metal enrichment, mixing and depletion. These provide interesting insights on the galactic thermal and chemical evolution, and may affect the mass calculation of metals and baryons.
\subsection{Temperature}\label{temp}
\noindent The presence of a warm-hot phase of the Milky Way CGM with log T(K)$\approx 6$ has been known from X-ray absorption line studies \citep{Gupta2012,Nicastro2016a,Nicastro2016b,Gupta2017,Nevalainen2017,Gatuzz2018}. Focusing on the sightline toward 1ES\,1553+113, our measurement of the temperature (and hydrogen column density) of the warm-hot phase is consistent with that of \cite{Gatuzz2018}, who assumed a solar metallicity and solar chemical composition. Their three-temperature model is equivalent with \texttt{ISMabs} ans \texttt{PHASE1} in our model, containing the cold/warm ISM and the warm-hot CGM. 

The strong \nex line detected in our high SNR spectra allowed us to discover the hotter (T$_2$) component. Our observations by themselves cannot determine if it resides in the Galactic ISM, CGM, or in the local group medium. From the knowledge of the different ISM phases studied so far, we know that the Galactic ISM is dominated by neutral and mildly ionized gas, and the distribution of the warm-hot phase indicates a significant contribution from the CGM \citep{Nicastro2016a,Nicastro2016b, Gatuzz2018}. Extrapolating this idea to the higher temperature, the hot gas likely resides in the extended region. Nonetheless, we investigated whether dense structures such as supernovae remnants (SNR) or superbubbles made by expanding and merging SNR are responsible for the observed hot gas phase:
\\
    a) There is no evidence of an individual SNR along our sightline, but the local hot bubble (LHB) is present. With a density of $\approx 3.9\pm0.4 \times 10^{-3}$ cm$^{-3}$ and path length of $\approx 100$ pc \citep{Snowden2014}, the LHB column density is $1.4 \times 10^{18}$ cm$^{-2}$, smaller than the column density of the hot gas by order(s) of magnitude. Thus the LHB contribution to the observed column density is negligible.\\
    b) The sightline to 1ES\,1553+113 ($l=21.91^\circ, b=43.96^\circ$) passes close to the Fermi Bubble (FB; $|l|\leq20^\circ,|b|<50^\circ$), therefore we investigated the possibility that our observed hot gas is from the structures in or around the FB \citep{Su2010,Kataoka2018}. The temperature inside the FB is believed to be hot \citep[$\geqslant10^7$K,][]{Su2010}; however our sightline does not pass through the FB. The sightline to 1ES\,1553+113 passes through the North Polar Spur (NPS). The temperature of NPS is $0.25-0.29$ keV \citep{Kataoka2018}, which is significantly lower than the temperature of the hot phase. Secondly, using the emission measure (0.02--0.07 cm$^{-6}$ pc) and the line-of-sight width ($\approx 5$ kpc) of the NPS \citep{Kataoka2018}, the column density is $3$--$5.6\times 10^{19}$ cm$^{-2}$, which is lower than the column density of the hot gas by a factor of $\sim 5$-$10$. Assuming the density of the X-ray shell around FB to be n$_{FB} \approx $10$^{-3}$cm$^{-3}$ as derived by \cite{Miller2016}, the implied path length for the hot component is L=N$_{H,2}^Z$/n$_{FB}= 100^{+80}_{-67}$ kpc. This is much larger than the spatial extent of the X-ray shell around the FB. While the NPS and the X-ray shell must contribute to the observed column density, they are unlikely to be a primary contributor unless the hot gas is much denser than that obtained by \cite{Miller2016}.
\\
  Next, we investigated whether the hot gas is from the Local Group. Using the $T_X-M$ relation for galaxy clusters: $T_X \propto M^{2/3}$, the temperature of the local group is $\approx 10^{6.69-6.91}$ K, assuming the local group mass = 6.4$\times 10^{12}$ M$_\odot$ \citep{Peebles1990}. This is consistent with T$_2$. However, our sightline ($21.91^\circ,43.96^\circ$) is away from M31 and the centre of the Local Group. Therefore, the contribution of the Local Group to the hot phase, if any, \textcolor{black}{might not be} significant. \cite{Oppenheimer2018} found that the thermal feedback can buoyantly rise to the outer CGM of MW-like halos of M$_{200} \leqslant 10^{12}$ M$_\odot$, moving baryons beyond the virial radius (R$_{200}$) and extending the CGM out to at least 2$\times$R$_{200}$. \textcolor{black}{Therefore, if the observed hot gas is extended beyond R$_{200}$, it is just a matter of nomenclature whether it should be called the CGM or the Local Group gas.}  
  
With the above considerations, we argue that the newly detected hot phase is primarily from the CGM of the Galaxy. Hydrodynamic simulations that take into account multistage stellar feedback from the bulges of MW-type galaxies predict that the temperature in the CGM can be as hot as T$_2$ \citep{Tang2009}. Analytic models predict that the CGM can significantly deviate from thermal equilibrium due to mechanical feedback (ejection of low angular momentum material) or thermal feedback (heating of the central regions), resulting in super-virial temperature as high as T$_2$ in the inner 50 kpc of the halo \citep{Pezzulli2017}. The absence of a strong \nex line in a deep $\sim$3 Ms absorption study towards PKS 2155-304 \citep{Nevalainen2017} indicates that this hot phase might not be ubiquitous, and is likely anisotropic. On the other hand, there have been hints of such hot gas in the halo of Milky Way in emission \citep{Henley2013,Nakashima2018} away from the Galactic centre, showing that the hot gas may not necessarily be related to the nuclear activity. However, emission is dominated by denser regions, so the hot gas detected in emission is perhaps located closer to the disk of the Galaxy. The presence of the hot gas in absorption with a high H column density of $\approx$ 10$^{20}$ cm$^{-2}$ indicates that the hot gas can in fact be present in a low density extended medium. \textcolor{black}{Based on combined emission and absorption measurements towards the Galactic bulge, \cite{Hagihara2011} found that a two-T hot ISM was necessary to reproduce the observed spectra. However, the two temperatures they obtain ($T_1=1.7\pm0.2\times 10^6$ K and $T_2=3.9^{+0.4}_{-0.3}\times 10^6$ K) differ by only a factor of 2, and their hotter component is significantly cooler than the hot gas we find. Thus the $\approx 10^7$ K hot gas we have discovered was not observed before either from the ISM or from the CGM of the Milky Way.} 
\subsection{Abundances}
We have measured two sets of abundance ratios: (1) oxygen, neon and nitrogen relative to iron; and (2) neon and nitrogen relative to oxygen (with three out of five of these being independent measurements). We find super-solar [Ne/Fe] and [O/Fe]; this shows that the CGM is $\alpha$-enhanced, indicating core-collapse supernovae enrichment \citep{Nomoto2006}. We also find super-solar [Ne/O] and similar [N/O]. The super-solar [Ne/O] is, at least partially, a result of the core-collapse supernovae enrichment \citep{Nomoto2006}. The large amount of nitrogen relative to neon suggests that there is an additional contribution to nitrogen, such as from dying asymptotic giant branch (AGB) stars \citep{Herwig2005}. However, it is highly unlikely that the two different sources of enrichment have contributed N and Ne to yield similar abundance ratios relative to oxygen. It is more likely that oxygen is sub-solar with respect to N and Ne. The lower oxygen abundance can be a result of different processes. Dust of silicates and other forms containing oxygen can be formed 300-600 days after core-collapse supernovae explosions \citep{Todini2001}. Thus, oxygen can be depleted onto dust in the ISM \textcolor{black}{(as found by \cite{Pinto2013} in the high-resolution X-ray absorption spectra of Galactic low-mass X-ray binaries)} before the outflows from the stellar feedback/supernovae reach the halo and enrich it with metals. Alternatively, if the CGM is inhomogeneously mixed as suggested by simulations \citep{Ford2013,Ford2014} and observed in cooler phases of the CGM \citep{Schaye2007}, oxygen can cool to lower temperatures without affecting the temperature of the gas at which other elements are abundant. This is reasonable to expect since the emissivity of oxygen at the CGM temperatures ($>10^6$ K) is $\sim3-10$ times higher than the emissivity of nitrogen and neon \citep{Bertone2013}. Thus, it is possible that nitrogen and neon remain in the hot/warm-hot phases while oxygen ``prefers" to transit to the warm or cooler phases. Similar transition of oxygen has already been found in cosmological simulations \citep{Ford2013} and observations \citep{Nevalainen2017}. If cooled sufficiently, oxygen may eventually get depleted onto circumgalactic dust. It requires more than 10$^7$ yrs to evaporate in a low-density environment with n $\approx$ 10$^{-3}$ cm$^{-3}$, so a significant amount of metals may be held in circumgalactic dust \citep{Tielens1994}. This is consistent with observations that a larger fraction of CGM metals (42\%) are in solid phase compared to that in the ISM ($\sim$30\%) \citep{Peeples2014}. While the deficit of oxygen with respect to neon and nitrogen are qualitatively consistent with the chemical enrichment by core-collapse supernovae and AGB stars, oxygen's depletion onto dust is likely to be necessary to explain the abundance ratios quantitatively. Determining the relative importance of all sources affecting the abundance ratios is beyond the scope of this paper. An alternative explanation might be as follows. Inhomogeneous mixing causes the density distribution of all ions (i.e. different ionization states of same/different elements) to be not necessarily the same \citep{Ford2013,Ford2014}. In that case, local abundance ratios,  which we measure along our line of sight, would  differ from the global average.

The hot gas (T$_2$) we discover in the Milky Way CGM along this sightline, as well as the non-solar abundance ratios, have not been detected in \textcolor{black}{the CGM of} any Milky Way-type external galaxies \citep{Strickland2004,Yamasaki2009}, \textcolor{black}{nor in most of the absorption studies of the Galactic ISM \citep{Yao2006a,Yao2006b,Yao2009b}. Mildly super-solar Ne/O has been observed in absorption in the warm-hot ISM \citep{Yao2009a}}, and $\alpha$-enhancement has been observed \textcolor{black}{in the CGM} in a couple of emission studies \citep{Strickland2004,Yamasaki2009,Nakashima2018}, though with poor constraints. In external galaxies, the $\alpha$-enhancement has been observed along the minor axes and in extra-planar regions within 10 kpc of the galactic disks \citep{Strickland2004,Yamasaki2009}; therefore it is associated with outflow/galactic fountain. As discussed in \S\ref{temp}, the hot gas we discover is spatially extended, and is primarily from the Milky Way CGM. Therefore, the $\alpha$-enhancement in the hot gas (figure \ref{fig:nonsolarfe}) shows that the feedback from the core collapse supernovae has extended to a large length scale. 
\subsection{Missing metals and missing baryons}
\noindent We have detected the hot component and non-solar abundance ratios in the Milky Way CGM along one sightline. Depending on the path-length, density and covering factor, it would affect the calculation of the baryonic mass and the metal mass of the CGM. 

The previous X-ray absorption-line studies were not sensitive to the non-solar abundance patterns, and the mass of metals in the warm-hot CGM was derived from the oxygen measurements:
\begin{equation}
    M(Z) = \frac{M(\hbox{O})}{f_O} 
\end{equation}
\begin{equation}
    M(\hbox{O})\propto N(\hbox{O})= \frac{N(\hbox{OVII})}{f_{OVII(T)}} \hbox{or,} \,\, \frac{N(\hbox{OVIII})}{f_{OVIII(T)}}
\end{equation}
here M(Z) and M(O) are respectively the total mass of metals and oxygen in the X-ray traced CGM and $f_O$ is oxygen-to-metal ratio. The temperature in the warm-hot (T $\approx 10^6$ K) phase was determined using the \oviiin/\ovii ratio. If a good fraction of \oviii arises in the hot (T $\approx 10^7$ K) phase, in a single-temperature model the \oviii ionization fraction would be over-estimated, and hence the warm-hot phase temperature ($T_1$) would be over-estimated. Depending on whether \ovii or \oviii is used to calculate the total mass of oxygen, the overestimation of temperature will lead to an over/underestimation of oxygen mass in the warm-hot phase, because $f_{OVII}$ decreases and $f_{OVIII}$ increases with temperature above 10$^6$ K. Inclusion of the hot component not only resolves this issue, it also traces an extra component of metals.  Secondly, sub-solar gas-phase oxygen, with the $f_O$ range of  $\approx$9\% to 28\%, compared to $f_O\approx 44\%$ for solar oxygen, may affect M(Z) by a factor of $\approx$2--5. This shows how the census of other metals, along with oxygen, in the X-ray traced CGM  is so vital.

The calculation of total baryonic mass 
\begin{equation}
    M_b=\frac{M(Z)}{0.02}\Big(\frac{Z}{Z_\odot}\Big)^{-1}
\end{equation} follows directly from the total metallic mass. For an assumed metallicity, if the oxygen is sub-solar as we find here, the total baryonic mass of the warm-hot CGM would be underestimated by the same factor (2-5 in our case) as the total metallic mass. Additionally, the hot CGM was not included in the previous works, so the mass in the X-ray traced CGM was underestimated. Once again, the contribution of the hot component to the baryonic budget would depend on its path-length, density, covering factor and volume-filling factor.

\subsection{Assumptions and Caveats}
\noindent We calculated relative abundances of metals using the solar model of \cite{Asplund2009}. If we use other values of relative solar abundances, the results remain the same: [O/Fe]= $0.8^{+0.7}_{-0.3}$, [Ne/O] = [N/O] = $0.7^{+1.6}_{-0.2}$ for \cite{Wilms2000} abundances, [O/Fe]= $0.8^{+0.7}_{-0.3}$, [Ne/O] = $0.8^{+1.6}_{-0.2}$ and [N/O] = $0.7^{+1.6}_{-0.2}$ for \cite{Lodders2003} abundances. This shows that our abundance measurement is not affected by the composition prescription. 

Our temperature estimates are subject to the assumption that the gas is in CIE. However, the gas at $T_2$ is not at the virial temperature of Milky Way, so it might not be in equilibrium.  If the hot gas is stretched across a large length scale ($\sim$ a few 100 kpc), the average density will be low enough to have the post-shock electron-ion relaxation timescale comparable with/longer than the Hubble time \citep{Yoshida2005}. In that case, the electron temperature will be overestimated and the abundance of metals will be over/underestimated. Similarly, if the gas in our sightline is photoionized (for reasons unknown) instead of being in CIE, the abundance ratios would be over/underestimated. 

As the cooling efficiency and dust depletion rate of N, O and Ne are different, the chemical composition in two temperature phases are not necessarily the same. However, we do not have any diagnostics of nitrogen lines or \ovii lines in the hotter phase, thus we cannot determine the abundance ratios in the two temperature phases independently, and we have assumed the ratios to be the same. The strongest constraints on iron abundance are for the hot phase, so in the least, our observed supersolar [Ne/Fe] and [O/Fe] are for the hot component. Similarly, the  observed super-solar [N/O] and [Ne/O] are, in the least, for the warm-hot component. Same abundance ratios are implicitly assumed in the derivation of the total column density of the hot phase N$_{H,2}^Z$, which may be under/overestimated. Also, the calculation of N$_{H,2}^Z$ involves the ionization fraction of individual lines. If the hot component is not at ionization equilibrium, the ionization fractions of H-like ions can be different from those at CIE \citep{Oppenheimer2013}. This will affect the derivation of N$_{H,2}^Z$. 

In the Discussions section, we have commented on how sub-solar oxygen may affect the calculations of metals and baryons in the CGM. A more complete calculation will have to involve the non-solar abundance ratios with respect to iron, and the abundance ratio of other $\alpha$-elements (C, Mg, Si) as well.

\section{Conclusion}\label{sec:summary}
\noindent In this paper, we have studied the z=0 absorber(s) in a deep X-ray grating spectra towards 1ES\,1553+113. Our work presents three interesting results:
\begin{enumerate}
    \item A hot $10^7$ K gas phase coexists with the warm-hot 10$^6$ K CGM, along this sightline
    \item The hot CGM shows significant $\alpha-$enhancement, likely due to core-collapse supernovae enrichment
    \item The abundance ratio of nitrogen, oxygen and neon in the warm-hot/hot CGM is significantly non-solar. Nitrogen and neon are in solar mixture, oxygen is in deficit with respect to nitrogen and neon. Along with the enrichment by AGB stars and core-collapse supernovae, it has possible hints of oxygen-depletion.
\end{enumerate}
These results provide insights on the thermal history, chemical enrichment and mixing in the circumgalactic medium, and provide important inputs to theoretical models of galaxy formation and evolution. 

It is necessary to extend such deep X-ray absorption analysis to many other sightlines to search for and characterize the temperature and chemical composition of the highly ionized CGM, using multiple tracer elements like carbon, nitrogen, neon, magnesium and silicon along with oxygen. At present, the archival data of \chandra and \xmm can be very useful in this regard. On a longer timescale, planned missions like \textit{XRISM, Athena, Lynx} in the next decade and beyond will offer an outstanding opportunity to observe the highly ionized diffused medium in unprecedented detail. This will bring us closer to understanding the co-evolution of the galaxy and its CGM.

\section*{Acknowledgements}
\noindent This work is based on observations obtained with \textit{XMM-Newton}, an ESA science mission with instruments and contributions directly funded by ESA Member States and NASA. S.M. acknowledges NASA grant NNX16AF49G. F.N. acknowledges funding from the INAF PRIN-SKA 2017 program 1.05.01.88.04. Y.K. acknowledges support from DGAPAPAIIPIT grant IN106518. 
\facilities{\xmm}
\software{HeaSoft v6.17 \citep{Drake2005}, NumPy v1.11.0 \citep{Dubois1996}, Matplotlib v1.5.3 \citep{Hunter2007}}


\end{document}